\documentclass[journal]{IEEEtran}
\ifCLASSINFOpdf
\else
\fi
\usepackage{amsmath}
\usepackage{amssymb}
\usepackage{autobreak}
\usepackage{amsfonts}
\usepackage{bm}
\usepackage{graphicx}
\usepackage{algorithm}  
\usepackage{algorithmicx}  
\usepackage{algpseudocode}
\usepackage{cite}
\usepackage{hyperref}
\usepackage{mathrsfs}
\usepackage{relsize}

\newenvironment{lequation}{
    \begin{equation}\large }{\end{equation}
}
\begin{document}
\setlength{\lineskip}{1.0pt}
\setlength{\lineskiplimit}{1.0pt}
%
\title{Weighted Sum Age of Information Minimization in  Wireless Networks with Aerial IRS}
%
%

\author{Wanting Lyu,~Yue Xiu,~Songjie Yang,~Phee Lep Yeoh,~\IEEEmembership{Member,~IEEE},~Yonghui Li,~\IEEEmembership{Fellow,~IEEE}, ~~~~~~~~~Zhongpei Zhang,~\IEEEmembership{Member,~IEEE} \\
        
        \thanks{Wanting Lyu, Songjie Yang and Zhongpei Zhang are with National Key Laboratory of Science and Technology on Communications, University of Electronic Science and Technology of China, Chengdu61173, China (E-mail: lyuwanting@yeah.net; yangsongjie@std.uestc.edu.cn; zhangzp@uestc.edu.cn).
        
        Yue Xiu is with Chengdu Research \& Development Centre, Huawei Technologies Co., Ltd., People's Republic of China (E-mail: xiuyuw12345678@163.com).
        
        Phee Lep Yeoh and Yonghui Li are with the School of Electrical and Information Engineering, University of Sydney, Sydney, NSW 2006, Australia (e-mail: phee.yeoh@sydney.edu.au; yonghui.li@sydney.edu.au).

        }
}

\maketitle

\begin{abstract}
In this letter, we analyze a terrestrial wireless communication network assisted by an aerial intelligent reflecting surface (IRS). We consider a packet scheduling problem at the ground base station (BS) aimed at improving the information freshness by selecting packets based on their AoI. To further improve the communication quality, the trajectory of the unmanned aerial vehicle (UAV) which carries the IRS is optimized with joint active and passive beamforming design. To solve the formulated non-convex problem, we propose an iterative alternating optimization problem based on a successive convex approximation (SCA) algorithm. The simulation results shows significant performance improvement in terms of weighted sum AoI, and the SCA solution converges quickly with low computational complexity.

\end{abstract}

\begin{IEEEkeywords}
Age of information, intelligent reflecting surface, unmanned aerial vehicle, scheduling, trajectory and beamforming design.
\end{IEEEkeywords}

%
\IEEEpeerreviewmaketitle

\section{Introduction}

\IEEEPARstart{I}{ntelligent} reflecting surface (IRS) has emerged as a promising technology to meet the ultra high demands on communication quality in beyond fifth-generation (B5G) mobile communications \cite{IRSmagzine}. Consisting of a large number of low-cost passive reflecting elements, IRS can reconfigure the wireless signal propagation environment to mitigate the channel impairments with high energy efficiency, which thus enhance the wireless links to improve the data rate and reliability \cite{IRStutorial}.


Existing works have explored the applications of IRS in wireless networks. Authors in \cite{YCLiang} studied the weighted sum-rate maximization through jointly optimizing the active and passive beamforming with perfect and imperfect channel state information (CSI). In \cite{SecretKey}, the secrete key generation capacity was improved with the assist of IRS. In \cite{LYHcellfree}, multiple IRSs cooperation was investigated to enhance energy efficiency in cell free MIMO networks. To improve the deployment flexibility of the IRS, an aerial IRS was proposed in \cite{AIRS_YZeng}, where the IRS is carried by an unmanned aerial vehicle (UAV). It is shown that aerial IRS can significantly increase the probability of a line-of-sight (LoS) link between the IRS and the ground nodes \cite{skyUAVcomm}.

Recently, real-time communications is becoming especially crucial in mission-critical application scenarios, such as vehicle-to-vehicle (V2V) networks and industrial Internet of things (IIoT). In these networks, data freshness is a key requirement to avoid accidents and errors caused by delayed information. To quantify the freshness of information, the authors of \cite{realtime} first introduced the
concept of the age of information (AoI), which is defined as the time past since the generation of the last successfully delivered and decoded information packet \cite{AoIswipt}. The authors in \cite{2018Age} analyzed and derived a closed-form expression for the average AoI using queuing theory and probability channel fading models. Considering that the relay node is undesirable to keep transmitting packets in real cases, in \cite{LYH_aoi}, an optimal scheduling policy was studied to minimize the AoI with a tradeoff of forwarding and receiving operation at the relay. In \cite{UAVAoI}, a UAV sensing and transmission time tradeoff design during a given period was proposed for a cellular Internet of UAVs to minimize the AoI.

In this paper, we use aerial IRS to improve the AoI in wireless communication networks. Our proposed solution is suitable for deployment in urban areas with densely distributed buildings, vehicles and human bodies, where the direct link between the terrestrial base station (BS) and user is likely to experience severe fading with non-line-of-sight (NLoS) channels. With the assistance of the UAV, the IRS has a higher probability to establish a LoS channel, and simultaneously, the deployment flexibility is dramatically increased to cater for real-time user demands. To analyse this network, we build a 3D environment model, and formulate a joint packet scheduling, active and passive IRS beamforming, and UAV trajectory optimization problem to minimize the total weighted sum AoI. An alternating optimization algorithm is developed to decouple the variables, with introduced slack variables and successive convex approximation (SCA) algorithm to deal with the non-convex constraints. Significant performance improvement measured by the sum AoI is verified by the numerical experiments, which also highlights the rapid convergence performance with low computational complexity.

\vspace*{-0.3\baselineskip}
\section{System Model and Problem Formulation}

\subsection{Wireless Communication Model}
\vspace*{-0.2\baselineskip}
\begin{figure}[!t]
    \centering
    \includegraphics[width=0.65\linewidth]{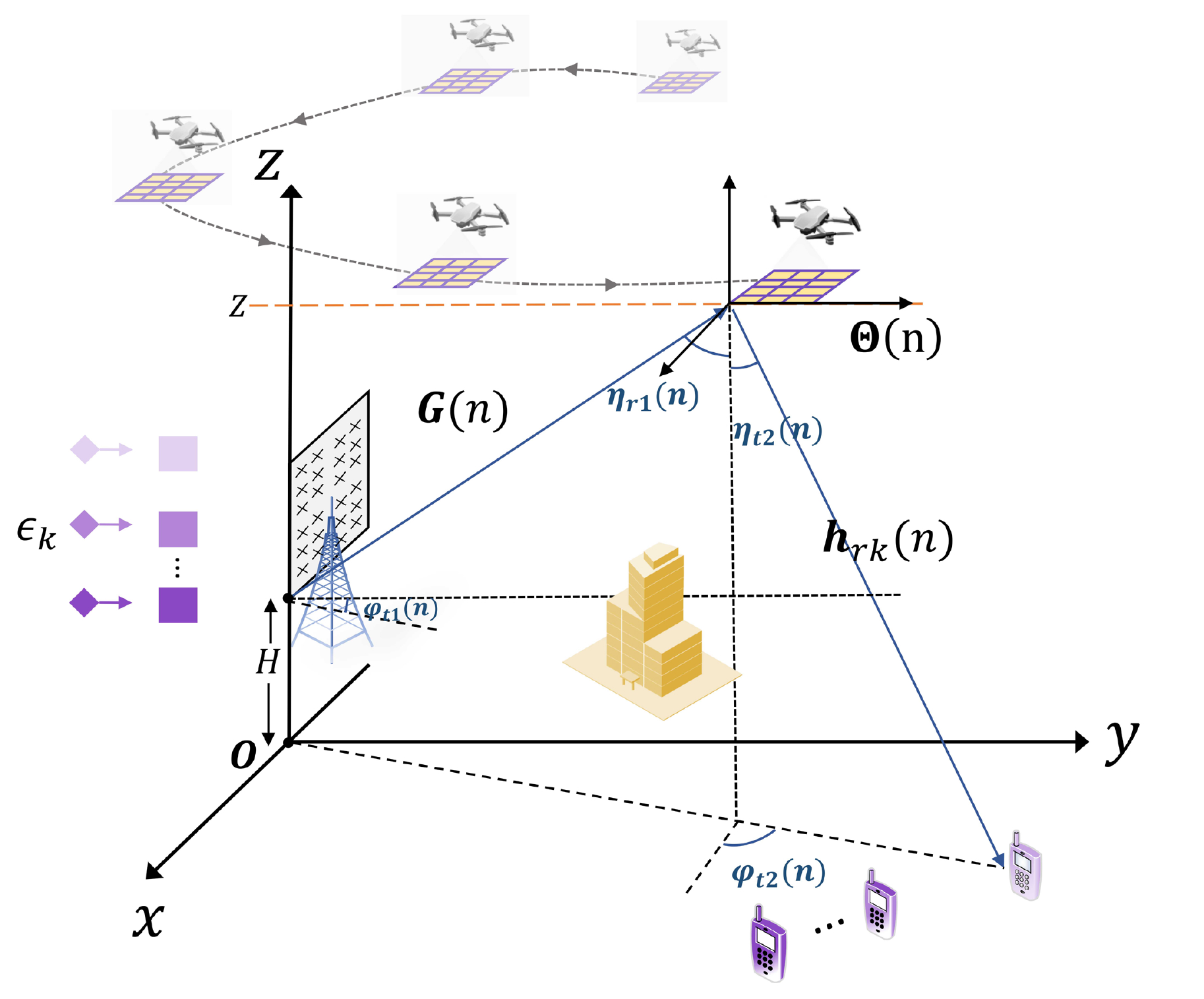}
    \caption{System model of the aerial IRS assisted network.}
    \label{model}
    \vspace*{-0.5\baselineskip}
\end{figure}

An aerial IRS aided wireless downlink system is considered as shown in Fig.1, where the IRS is hovering above the ground carried by a UAV. The ground base station (BS) is equipped with a uniform planar array (UPA) with $M = M_xM_z$ antennas serving $K$ single-antenna ground users. The IRS is also configured with a UPA with $N_s = N_{sx}N_{sy}$ reflecting elements. The time of the whole aerial IRS-assisted communications ($T$ for total) is divided into $N_T$ time slots each with duration $\Delta = T/N_T$, where $n$ is used to denote the current time slot.

The BS with height $H$ is located at the origin of the coordinate system, and the antenna array is located on the x-z plane. The space between adjacent antennas is $d_{ox} = d_{oz} = \frac{\lambda}{2}$, where $\lambda$ is the wavelength. The first antenna at the BS is assumed to be located at $\bm{c} = [0,0,H]^T$. Accordingly, the coordinate of the $m_x^{th}$ column, $m_z^{th}$ row antenna is $\bar{\bm{c}} = \bm{c} + [0, (m_x-1)d_{ox}, (m_z-1)d_{oz} ], \forall m_x \in \{1,2,...,M_x\}, m_z \in \{1,2,...,M_z\}$. The IRS is placed on the $x-y$ plane with a fixed altitude of $Z$. We consider the first reflecting element to be at the corner of the IRS, with coordinate $\bm{q}(n) = [q_x(n), q_y(n), Z]^T$ at time slot $n$. Fixed distances $d_x < \frac{\lambda}{2}$ and $d_y < \frac{\lambda}{2}$ are set for separating the adjacent elements \cite{AIRS_YZeng}. Similarly, the location of the $n_s^{th}$ IRS element is denoted as $\bar{\bm{q}}(n) = [q_x(n) + (n_{sx}-1)d_x, q_y(n) + (n_{sy}-1)d_y, q_z(n)]^T, \forall n_{sx} \in \{1,2,...,N_{sx}\}, n_{sy} \in \{ 1,2, ... ,N_{sy}\}$. The users are assumed to be on the ground, and the coordinate of user $k$ is $\bm{l}_k = [k_x, k_y, 0]^T, \forall k \in \{1,2,...,K\}$. 

The distance between the reference points at the transmit antenna array and the IRS is thus $d_{br}(n) = ||\bm{q}(n) - \bm{c}||_2$, and the distance between the first reflecting element and the $k^{th}$ user is $d_{rk}(n) = ||\bm{q}(n) - \bm{l}_k||_2$. The UAV flight platform is sufficiently high so that the channels between the aerial IRS and the ground nodes are mainly LoS links with a high probability \cite{skyUAVcomm}. Due to the dense urban environment, assume that the direct links from the BS to the users are blocked. Without loss of generality, the free space propagation model is considered for measuring the large-scale attenuation. Therefore, the path loss of the BS-IRS channel and the IRS-user $k$ channel are $\rho_{br}(n) = \rho_o\frac{d_o^2}{d_{br}^2(n)}$ and $\rho_{rk}(n) = \rho_o\frac{d_o^2}{d_{rk}^2(n)}$ respectively, where $\rho_o$ is the reference path loss at a given reference distance $d_o = 1 \rm m$.

It is noteworthy that the IRS can be approximated as a UPA since its size is sufficiently small compared with the distance between the communication links. As shown in Fig.2, the angle of departure (AoD) at the BS and the angle of arrival (AoA) are approximately symmetric.



Accordingly, the receiving array response vector at the IRS at time slot $n$ is 
\vspace{-2mm}
\begin{align}
    \bm{a}_{R1} &(\eta_{r1}(n),\varphi_{r1}(n)) \nonumber \\
    & = [ 1,\; e^{-j\phi_{r1x}(n)} ,\;...,\; e^{-j\phi_{r1x}(n)(N_{sx}-1)} ]^T \nonumber \\
    &\otimes [ 1,\; e^{-j\phi_{r1y}(n)} ,\;...,\; e^{-j\phi_{r1y}(n)(N_{sy}-1)} ]^T \in \mathbb{C}^{N_s \times 1},
    \label{a_R1}
\end{align}
\noindent where $\phi_{r1x}(n) = \frac{2\pi}{\lambda}d_x \sin\eta_{r1}(n) \cos\varphi_{r1}(n)$ and $\phi_{r1y}(n) = \frac{2\pi}{\lambda}d_y \sin\eta_{r1}(n) \sin\varphi_{r1}(n)$. The variables $\eta_{r1}(n)$ and $\varphi_{r1}(n)$ denote the elevation and azimuth AoA at time slot $n$ at the IRS, and $\cos(\eta_{r1}(n)) = \frac{Z-H}{||\bm{q}-\bm{c}||_2}$, $\sin(\varphi_{r1}(n)) = \frac{q_y}{||\bm{q}-H||_2}$ .  

The transmitting array response vector $\bm a_{T1}(n)$ at the BS antenna array is defined similar to (\ref{a_R1}). Hence the link between the BS and the IRS is defined as 
\vspace*{-0.2\baselineskip}
\begin{align}
    \bm{G}&(n) =  \sqrt{\rho_{br}(n)} e^{-j\frac{2\pi d_{br}(n)}{\lambda}} \bm{a}_{R1}(\eta_{r1}(n), \varphi_{r1}(n)) \nonumber \\ 
    & \bm{a}_{T1}^H(\eta_{t1}(n),\varphi_{t1}(n)).
    \label{chann_G}
\end{align}
The IRS-user link is expressed as
\vspace*{-0.3\baselineskip}
\begin{equation}
    \bm{h}_{rk}^H(n) = \sqrt{\rho_{rk}(n)} e^{-j\frac{2\pi d_{rk}(n)}{\lambda}}  \bm{a}_{T2}^H(\eta_{t2}(n),\varphi_{t2}(n)),
    \label{chann_hk}
\end{equation}
\noindent where the transmitting array response vector at the IRS is 
\vspace*{-0.5\baselineskip}
\begin{align}
    \bm{a}_{T2} &(\eta_{t2}(n),\varphi_{t2}(n)) \nonumber \\
    & = [ 1,\; e^{-j\phi_{t2x}(n)} ,\;...,\; e^{-j\phi_{t2x}(n)(N_{sx}-1)} ]^T \nonumber \\
    &\otimes [ 1,\; e^{-j\phi_{t2y}(n)} ,\;...,\; e^{-j\phi_{t2y}(n)(N_{sy}-1)} ]^T \in \mathbb{C}^{N_s \times 1},
    \label{a_T2}
\end{align}
and $\phi_{t2x}(n) = \frac{2\pi}{\lambda}d_x \sin\eta_{t2}(n) \cos\varphi_{t2}(n)$ and $\phi_{t2y}(n) = \frac{2\pi}{\lambda}d_y \sin\eta_{t2}(n) \sin\varphi_{t2}(n)$. 

The phase shift matrix at the IRS is represented as $\bm{\Theta}(n) = {\rm diag}\{e^{j\theta_1(n)},\; e^{j\theta_2(n)},...,\; e^{j\theta_{N_s}(n)} \}$, where $\bm{\theta}(n) = [\theta_1(n), \theta_2(n),...,\theta_{Ns}(n)]^T$. $\theta_{n_s}(n) \in [0,2\pi)$ is the $n_s^{th}$ phase shift of the IRS for $n_s\in\{1,...,N_s\}$.
The transmit symbol for user $k$ is $s_k(n)$ satisfying $\mathbb{E}\{|s_k(n)|^2\} = 1$, and $\bm{w}_k(n) \in \mathbb{C}^{M\times 1}$ is the corresponding transmit beamforming vector. As such, the transmitted signal at the BS can be expressed as 
\vspace*{-0.5\baselineskip}
\begin{equation}
    \bm{x}(n) = \sum_{k=1}^K \bm{w}_k(n) s_k(n),
    \label{Tx_sig}
\end{equation}
\noindent where we set the constraint that, the transmit power for each user is less than or equal to $P_o$,
\vspace*{-0.5\baselineskip}
\begin{equation}
    ||\bm{w}_k(n)||^2 \le P_o, \; \forall k \in \{1,...,K\}.
    \label{Cons_pow}
\end{equation}

Assuming the direct links are unavailable, the received signal at user $k$ can be expressed as 
\vspace*{-0.5\baselineskip}
\begin{equation}
    y_k(n) = \bm{h}_{rk}^H(n)\bm{\Theta}(n)\bm{G}(n)\bm{x}_n + u_k(n),
    \label{Rx_sig}
\end{equation}
\noindent where $u_k(n) \sim \mathcal{CN}(0,\sigma_o^2)$ represents the additive white Gaussian noise (AWGN) at the receiver of user $k$. For simplicity, the channel for user $k$ can be denoted as $\bm{h}_k^H(n) = \bm{h}_{rk}^H(n)\bm{\Theta}(n)\bm{G}(n)$, and $y_k(n) = \bm{h}_k^H(n)\bm{x}(n) + u_k(n)$. 

\subsection{The Age of Information Model}

In this model, we assume $K$ data streams at the BS corresponding to $K$ users, where a status update data packet of stream $k$ arrives at the base station with a probability $\epsilon_k$ at each time slot, and we define $p_k(n) \in \{0,1\}$ to indicate whether there is a packet arriving, with $Pr[p_k(n) = 1] = \epsilon_k$. $A_k(n)$ is used to denote the AoI of the targeted user $k$ at time slot $n$. We use a binary scheduling indicator $\alpha_k(n)=1$ when the $k^{th}$ data stream is scheduled otherwise it is equal to $0$. We furthur assume that there is only one available channel at each time slot, such that
\vspace*{-0.5\baselineskip}
\begin{equation}
    \alpha_k(n) \in \{0,1\}, \; \forall n.
    \label{Cons_alpha}
\end{equation}
\vspace*{-0.7\baselineskip}
\begin{equation}
    \sum_{k=1}^K \alpha_k(n) \le 1,\; \forall n.
    \label{Cons_channel}
    \vspace*{-0.3\baselineskip}
\end{equation}
The AoI will increase linearly without newly updated data packets successfully received by the targeted receiver, and otherwise it is determined by the packet queuing time $z_k(n)$, with $A_k(n+1) = z_k(n)+1$. Following \cite{UAVAoI} and \cite{RISAoI}, the system time $z_k(n)$ is given by
\vspace*{-0.2\baselineskip}
\begin{equation}
    z_k(n+1) = \left \{
    \begin{aligned}
        & 0,      & if\, \alpha_k(n+1) = 1, \forall k,n \\
        & z_k(n) + 1, & otherwise.
    \end{aligned}
    \right.
    \label{systime}
    \vspace*{-0.2\baselineskip}
\end{equation}

We define another binary variable $\xi_k(n)$ indicating the available packet status with the value equalling to $1$ when there is a available packet at stream $k$. Otherwise $\xi_k(n)$ only turns to $0$ with a successful delivery at last time slot and no new packet. A packet is delivered successfully if the scheduled stream has an available packet and the received signal-to-noise ratio (SNR) is no less than the threshold value. Therefore the indicator $\xi_k(n)$ can be expressed as
\begin{equation}
    \xi_k(n+1) = p_k(n+1) + \xi_k(n)(1-\alpha_k(n))(1-p_k(n+1)),
    \label{xi_re}
\end{equation}
\noindent where $\gamma_{th}$ denotes the threshold SNR. As a result, the AoI can be expressed as 
\vspace*{-0.2\baselineskip}
\begin{align}
    &A_k(n+1)  =\,  z_k(n)\alpha_k(n)\xi_k(n) + A_k(n)[(1-\alpha_k(n))\xi_k(n)  \nonumber \\ & + \alpha_k(n)(1-\xi_k(n)) + (1-\alpha_k(n))(1-\xi_k(n)) ] + 1,
    \label{AoI}
    \vspace*{-0.5\baselineskip}
\end{align}
\noindent with the SNR constraint
\vspace*{-0.2\baselineskip}
\begin{equation}
    \gamma_k(n) \ge \alpha_k(n)\xi_k(n)\gamma_{th},
    \label{Cons_SINR}
    \vspace*{-0.2\baselineskip}
\end{equation}
\noindent where the received SNR is
\vspace*{-0.2\baselineskip}
\begin{equation}
    \gamma_k(n) = \frac{|\bm{h}_k^H(n)\bm{w}_k(n)|^2}{\sigma_o^2}.
    \label{SNR_k}
    \vspace*{-0.5\baselineskip}
\end{equation}

\subsection{Problem Formulation}

To meet the demands of real-time communications and guarantee the freshness of information, the aim of this paper is to minimize the weighted sum AoI of all $K$ users. We propose an algorithm to optimize the UAV trajectory, scheduling scheme at the BS and active and passive beamforming design. The optimization problem is formulated as (P1),
\vspace*{-0.5\baselineskip}
\begin{align}
    \mathrm{(P1)}: \; & \min_{\bm{q}(n),\alpha_k(n),\atop\bm{w}_k(n),\bm{\Theta}(n)} \sum_{k=1}^K b_kA_k^{N_T} \label{P1obj}\\
    \bm{s.t.} \quad & (\ref{Cons_pow}),(\ref{Cons_alpha}), (\ref{Cons_channel}), (\ref{Cons_SINR}) \nonumber \\
    & q_z(n) = Z, \; \forall n \in \{1,...,N_T\}, \label{Cons_UAValti}\tag{\ref{P1obj}{a}} \\
    & ||\bm{q}(n) - \bm{q}(n-1)|| \le v_{max}\Delta, \label{Cons_velocity} \tag{\ref{P1obj}{b}}\\
    & \theta_{n_s}(n) \in [0,2\pi)
    ,\forall n_s \in\{1,...,N_s\}, \label{Cons_phase}\tag{\ref{P1obj}{c}}
    \vspace*{-0.5\baselineskip}
\end{align}
\noindent where $b_k$ is the priority for user $k$, and $A_k^{N_T}$ denotes the sum AoI of user $k$ for $N_T$ total time slots.

Since the AoI will keep increasing linearly without newly updated data packets, the problem of minimizing sum AoI becomes maximizing the AoI reduction at each time slot. Specifically, the stream with the worst weighted AoI should be scheduled if there is an available packet to be delivered. Hence the original problem (P1) can be reformulated as
\vspace*{-0.5\baselineskip}
\begin{align}
    \mathrm{(P2)}: \; & \max_{\bm{q}(n),\alpha_k(n), \atop\bm{w}_k(n),\bm{\Theta}(n)} \sum_{k=1}^K b_kA_k(n)\alpha_k(n)\xi_k(n) \label{P2obj}\\
    \bm{s.t.} \quad & (\ref{Cons_pow}),(\ref{Cons_alpha}), (\ref{Cons_channel}), (\ref{Cons_SINR}),(\ref{Cons_UAValti}),(\ref{Cons_velocity}),(\ref{Cons_phase}).      \nonumber 
    \vspace*{-0.5\baselineskip}
\end{align}

However, the problem is intractable because of i) the coupling of variables $\bm q(n),\,\alpha_k(n),\, \bm w_k(n)$ and $\bm \Theta(n)$; and ii) the non-convexity of constraints (\ref{Cons_alpha}) and (\ref{Cons_SINR}). The proposed iterative solution based on SCA algorithm is derived and elaborated as follows.

\vspace*{-0.4\baselineskip}

\section{Proposed Solutions}
\vspace*{-0.2\baselineskip}
In this section, the proposed algorithm is given as follows. To decouple the variables, the problem is solved alternatively, where the active and passive beamforming are jointly designed for maximizing the array gain with fixed UAV location, and given the optimized beamforming design, the scheduling policy is updated with UAV trajectory optimization. The SCA algorithm is an efficient approach to deal with the non-convexity of the AoI constraints.
\vspace*{-0.2\baselineskip}
\subsection{Joint Active and Passive Beamforming Design}

To maximize the AoI reduction at each time slot, the received SNR should be maximized to satisfy constraint (\ref{Cons_SINR}). Therefore, it is equivalent to maximizing the array gains of the UPAs at the BS and IRS respectively through digital beamforming and IRS phase shift design. 

Given the UAV location, the optimal transmit beamforming can be determined by Maximum Ratio Transmission (MRT) to align with the transmitting array response vector at the BS, which is given by 
\vspace*{-0.5\baselineskip}
\begin{equation}
    \bm{w}_k(n) = \sqrt{P_k}\frac{\bm{a}_{T1}(\eta_{t1}(n),\varphi_{t1}(n))}{\sqrt{M}}. 
    \label{opt_w}
    \vspace*{-0.5\baselineskip}
\end{equation}

The array gain at the IRS is expressed as
\vspace*{-0.2\baselineskip}
\begin{lequation}\displaystyle
    \Big|\sum\limits_{n_{sx}=1}^{N_{sx}} \sum\limits_{n_{sy}=1}^{N_{sy}}   e^{j\left( \theta_{n_{sx},n_{sy}}(n) + \mu_{n_{sx}}(n) + \mu_{n_{sy}}(n) \right)} \Big|,
    \label{array_gain}
    \vspace*{-0.5\baselineskip}
\end{lequation}
\noindent where $\mu_{n_{sx}}(n) = (n_{sx}-1)(\phi_{t2x}(n)-\phi_{r1x}(n))$ and $\mu_{n_{sy}}(n) = (n_{sy}-1)(\phi_{t2y}(n)-\phi_{r1y}(n))$.

Applying the triangle inequality,
\begin{large}
\begin{align}
    &\Big|\sum\limits_{n_{sx}=1}^{N_{sx}} \sum\limits_{n_{sy}=1}^{N_{sy}} e^{j\left( \theta_{n_{sx},n_{sy}}(n) + \mu_{n_{sx}}(n) + \mu_{n_{sy}}(n) \right)}\Big| \nonumber  \\
    \le &\big|e^{j\theta_{1,1}(n)}\big| +
    \dots + \big|e^{j\left( \theta_{n_{sx},n_{sy}}(n) + \mu_{n_{sx}}(n) + \mu_{n_{sy}}(n) \right)} \big| + \nonumber \\
    & \dots +\big|e^{j\left( \theta_{N_{sx},N_{sy}}(n) + \mu_{N_{sx}}(n) + \mu_{N_{sy}}(n) \right)} \big| = N_s,
    \label{triangle}
    \vspace*{-0.5\baselineskip}
\end{align}
\end{large}
\noindent where the equality holds with 
\begin{small}
\begin{align}
    &\theta_{n_{sx},n_{sy}}(n) = -(\mu_{n_{sx}}(n)+\mu_{n_{sy}}(n)) \nonumber\\
    = & -\frac{2\pi}{\lambda}\Big[ d_x(n_{sx}-1) (\sin\eta_{t2}(n)\cos\varphi_{t2}(n) - \sin\eta_{r1}(n)\cos\varphi_{r1}(n))    \nonumber  \\
    & - d_y(n_{sy}-1) (\sin\eta_{t2}(n)\sin\varphi_{t2}(n) -   \sin\eta_{r1}(n)\sin\varphi_{r1})(n) \Big].
    \label{opt_theta}
    \vspace*{-0.5\baselineskip}
\end{align}
\end{small}

\subsection{Scheduling and UAV Trajectory Optimization}

With the above optimal active and passive beamforming strategies, the received SNR at user $k$ can be obtained as 
\vspace*{-0.3\baselineskip}
\begin{small}
\begin{align}
    \bar{\gamma}_k(n) = & \frac{\rho_{rk}(n)\rho_{br}(n)P_o N_s^2 M}{\sigma_o^2} \nonumber \\
    & = \frac{ \rho_o^2 d_o^4P_o N_s^2  M }{d_{rk}^2(n)d_{br}^2(n)\sigma_o^2} \nonumber \\
    & = \frac{ \rho_o^2 d_o^4P_o N_s^2  M }{ ||\bm{q}(n)-\bm{l}_k||^2||\bm{q}(n)-\bm{c}||^2 \sigma_o^2},
    \label{SNR_bar}
\end{align}
\end{small}
\noindent which is only related to the UAV location $\bm{q}(n)$. Therefore, the problem (P2) can be reformulated as
\vspace*{-0.5\baselineskip}
\begin{small}
\begin{align}
    \mathrm{(P3)}: \; & \max_{\bm{q}(n),\alpha_k(n)} \sum_{k=1}^K b_kA_k(n)\alpha_k(n)\xi_k(n) \label{P3obj}\\
    \bm{s.t.} \quad & \frac{ \rho_o^2 d_o^4P_o N_s^2  M }{ ||\bm{q}(n)-\bm{l}_k||^2||\bm{q}(n)-\bm{c}||^2 \sigma_o^2} \ge \alpha_k(n)\xi_k(n)\gamma_{th}  \label{Cons_SNR1}\tag{\ref{P3obj}{a}} \\
    &(\ref{Cons_alpha}), (\ref{Cons_channel}),(\ref{Cons_UAValti}),(\ref{Cons_velocity}),     \nonumber 
\end{align}
\end{small}
\noindent where the objective function and the available channel constraint (\ref{Cons_channel}) are linear, and the scheduling indicator constraint (\ref{Cons_alpha}) can be relaxed as 
\vspace*{-0.5\baselineskip}
\begin{equation}
    \alpha_k(n) \in [0,1],
    \label{Cons_rela_alpha}
    \vspace*{-0.5\baselineskip}
\end{equation}
which is convex. The UAV altitude constraint (\ref{Cons_UAValti}) can be expressed as 
\vspace*{-0.5\baselineskip}
\begin{equation}
    \bm{q}^{\rm T}(n)\bm{\varepsilon} = Z,
    \label{Cons_UAValti_cvx}
    \vspace*{-0.5\baselineskip}
\end{equation}
where $\bm{\varepsilon} = [0,0,1]^T$.
However, problem (P3) is still non-convex due to the constraint (\ref{Cons_SNR1}). To tackle the non-convexity, the SCA algorithm is applied iteratively. First, slack variables $r_{rk}(n)$ and $r_{br}(n)$ are introduced as the upper bound of $d_{rk}^2(n)$ and $d_{br}^2(n)$ respectively, which can be described as
\vspace*{-0.5\baselineskip}
\begin{align}
    & ||\bm{q}(n)-\bm{l}_k||^2 \le r_{rk}(n), \label{Cons_slack1} \\
    & ||\bm{q}(n)-\bm{c}||^2 \le r_{br}(n). \label{Cons_slack2}
    \vspace*{-0.5\baselineskip}
\end{align}
\noindent Thus, the lower bound of the $\bar{\gamma}_k(n)$ can be expressed as
\begin{equation}
    \bar{\gamma}_k(n) \ge \frac{ \rho_o^2 d_o^4P_o N_s^2  M }{ r_{rk}(n)r_{br}(n)\sigma_o^2} = \frac{\Gamma_k(n)}{r_{rk}(n)r_{br}(n)},
    \label{Cons_slackSNR}
    \vspace*{-0.2\baselineskip}\end{equation}
where $\Gamma_k(n) = \frac{ \rho_o^2 d_o^4P_o N_s^2  M }{ \sigma_o^2}$, and the SNR constraint is obtained as 
\begin{small}
\begin{equation}
    \frac{\Gamma_k(n)}{r_{rk}(n)r_{br}(n)} \ge \alpha_k(n)\xi_k(n)\gamma_{th},
    \vspace*{-0.5\baselineskip}
\end{equation}
\end{small}
\noindent which is equivalent to 
\begin{small}
\begin{equation}
    \frac{\Gamma_k(n)}{\xi_k(n)\gamma_{th}} \ge r_{rk}(n)r_{br}(n)\alpha_k(n).
    \label{Cons_r}
\end{equation}
\end{small}

\noindent To tackle the coupling between the variables, the logarithm is taken at both sides. Then, we compute the first-order Taylor series of the right-hand side (RHS) in the $i^{th}$ iteration in (\ref{Cons_r}), which results in
\vspace*{-0.5\baselineskip}
\begin{align}
    & \ln{\Gamma_k(n)} - \ln\xi_k(n)\gamma_{th} \ge \ln r_{rk}(n) + \ln r_{br}(n) + \ln\alpha_k(n) \nonumber\\
    = & \ln{r_{rk}^{(i-1)}(n)} + \ln{r_{br}^{(i-1)}(n)}  + \ln{\alpha_k^{(i-1)}(n)} \nonumber \\  
    & + \frac{r_{rk}^{(i)}(n) - r_{rk}^{(i-1)}(n)}{r_{rk}^{(i-1)}(n)} + \frac{r_{br}^{(i)}(n) - r_{br}^{(i-1)}(n)}{r_{br}^{(i-1)}(n)} \nonumber \\ 
     &+ \frac{\alpha_k^{(i)}(n) - \alpha_k^{(i-1)}(n)}{\alpha_k^{(i-1)}(n)}  =  \hat{r}_{rk}^{(i)}(n) + \hat{r}_{br}^{(i)}(n) + \hat{\alpha_k}^{(i)}(n),
    \label{TayGamma}
\end{align}
where $(i-1)$ denotes the value in the $i-1^{th}$ iteration. It is noteworthy that $\alpha_k(n)$ cannot be equal to $0$ for validity, so that constraint (\ref{Cons_rela_alpha}) becomes
\vspace*{-0.5\baselineskip}
\begin{equation}
    \alpha_k(n) \in (0,1].
    \label{Cons_rela_alpha_0}
    \vspace*{-0.5\baselineskip}
\end{equation}
Hence the problem (P3) in the $i^{th}$ iteration becomes 
\vspace*{-0.5\baselineskip}
\begin{small}
\begin{align}
    &\mathrm{(P3.1)}: \;  \max_{\bm{q}(n),\alpha_k(n),\atop r_{rk}(n), r_{br}(n)} \sum_{k=1}^K b_kA_k(n)\alpha_k(n)\xi_k(n) \label{P3.1obj}\\
    \bm{s.t.}\;
    & \hat{r}_{rk}^{(i)}(n) + \hat{r}_{br}^{(i)}(n) + \hat{\alpha}_k^{(i)}(n) \le \ln{\Gamma_k(n)} - \ln \alpha_k(n)\xi_k(n)\gamma_{th}  \tag{\ref{P3.1obj}{a}} \\
    &(\ref{Cons_channel}),(\ref{Cons_velocity}),(\ref{Cons_rela_alpha_0}),(\ref{Cons_UAValti_cvx}),(\ref{Cons_slack1}),(\ref{Cons_slack2}) \nonumber, 
\end{align}
\end{small}
\noindent which is a convex problem and can be efficiently solved using standard optimization solver such as CVX. Finally, the SCA based algorithm is summarized in Algorithm 1.
\begin{algorithm}[t]  
  \caption{SCA based joint scheduling and UAV trajectory optimization}  
  \begin{algorithmic}[1]  
      \State \textbf{Initialize} $ \bm q(n),\alpha_k(n), r_{rk}(n), r_{br}(n) \; \forall k \in \{1,..,K\}$, $\forall n \in \{1,...,N_T\}.$ Set iteration index $i = 1$.
      \Repeat
       \State \textbf{Update} $\bm q(n),\alpha_k(n), r_{rk}(n), r_{br}(n)$  by solving problem (P3.1).
      \Until{Convergence.}
      \State \textbf{Output} $\bm q(n),\alpha_k(n)$.
  \end{algorithmic}  
\end{algorithm}

\vspace*{-0.3\baselineskip}

\section{Numerical Results}

In this section, the results of the numerical simulations are provided to demonstrate the performance improvement of the proposed algorithm. A base station with height $H = 25 {\rm\, m}$ is considered, with a $M = 4\times 4$ UPA. $K = 6$ stationary ground users are located at $\bm l_1 = [200, -100, 0]^{\rm T} {\rm m}$, $\bm{l}_2 = [150, 300, 0]^{\rm T} {\rm m}$, $\bm{l}_3 = [320,  -280, 0]^{\rm T} {\rm m}$, $\bm{l}_4 = [490, 20, 0]^{\rm T} {\rm m}$, $\bm{l}_5 = [50, -200, 0]^{\rm T} {\rm m}$, $\bm{l}_6 = [730, 30, 0]^{\rm T} {\rm m}$ respectively. The maximum transmit power for each user is set to be $P_o = 1 {\rm W}$, and the noise power is $\sigma_o^2 = -110\, {\rm dBm}$. According to the carrier frequency $f_c = 2.4\, {\rm GHz}$, the reference channel power gain is set to be $\rho_o = -40\, {\rm dB}$. Besides, the distance between the adjacent elements at the IRS is $d_x = d_y = \lambda/10$. The maximum flying velocity is set as $v_{max} = 5\,{\rm m/s}$. We consider $N_T = 40$ simulation time slots in total, with each slot lasting $\Delta = 0.1\,{\rm s}$. The initial location of the UAV is set as $\bm{q} = [0,0,Z]^{\rm T}$, where the flight altitude $Z$ is set as different values for comparison.

Fig.3 plots AoI for $N_T = 40$ time slots of the $6$ users. The height of the aerial IRS is fixed as $Z=100 \,{\rm m}$. The number of IRS elements is assumed to be $N_s = 550$, and the SNR threshold is set as $\gamma_{th} = 25 {\rm\, dB}$. It can be seen that the AoI would increase linearly without the corresponding stream being scheduled. Therefore, when the AoI is high, it will drop due to packet scheduling and and transmission by the base station.
\begin{figure}[!t]
    \centering
    \includegraphics[width=0.65\linewidth]{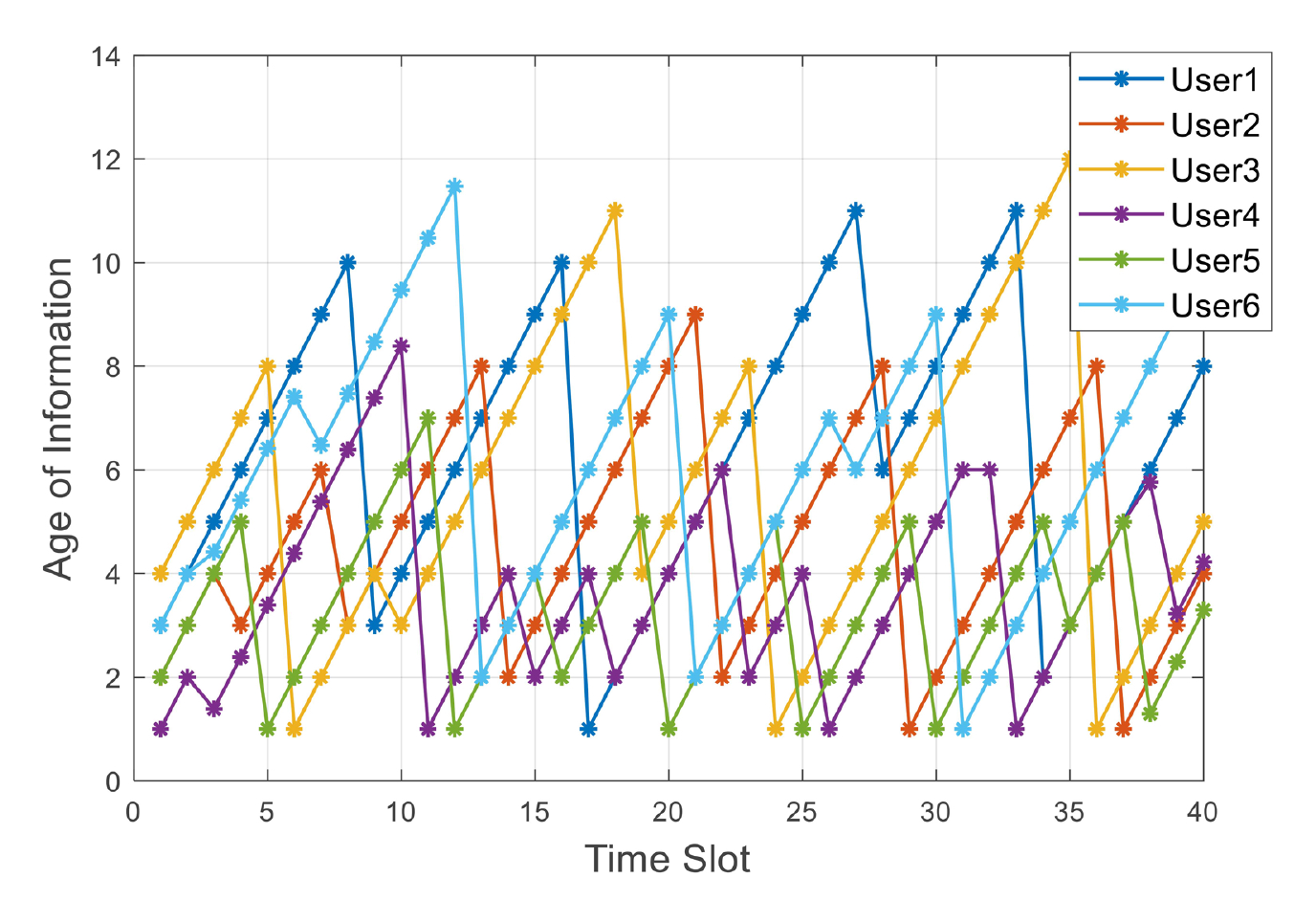}
    \caption{AoI performance of 6 users.}
    \label{AoIPerform}
\end{figure}

Fig.4 illustrates the weighted sum AoI for 40 time slots in total versus the number of IRS elements. The proposed algorithm shows significantly better performance than the fixed location scheme for the UAV at $[0, 0, 100\rm m]^{\rm T}$ in terms of information freshness. Simultaneously, the weighted sum AoI decreases with the increase of IRS elements $N_s$, from $150$ to $650$, converging to the optimal value. Thus, with larger-scale IRS, the communication link can be enhanced with the proposed algorithm. Besides, different levels of the threshold value of SNR are considered, namely $\gamma_{th} = 25 \,{\rm dB} ,30 \,{\rm dB},32\, {\rm dB}$ and $35\,{\rm dB}$ respectively,  and it can be seen that with higher $\gamma_{th}$, the freshness of the received packets becomes worse.  

\begin{figure}[!t]
    \centering
    \includegraphics[width=0.7\linewidth]{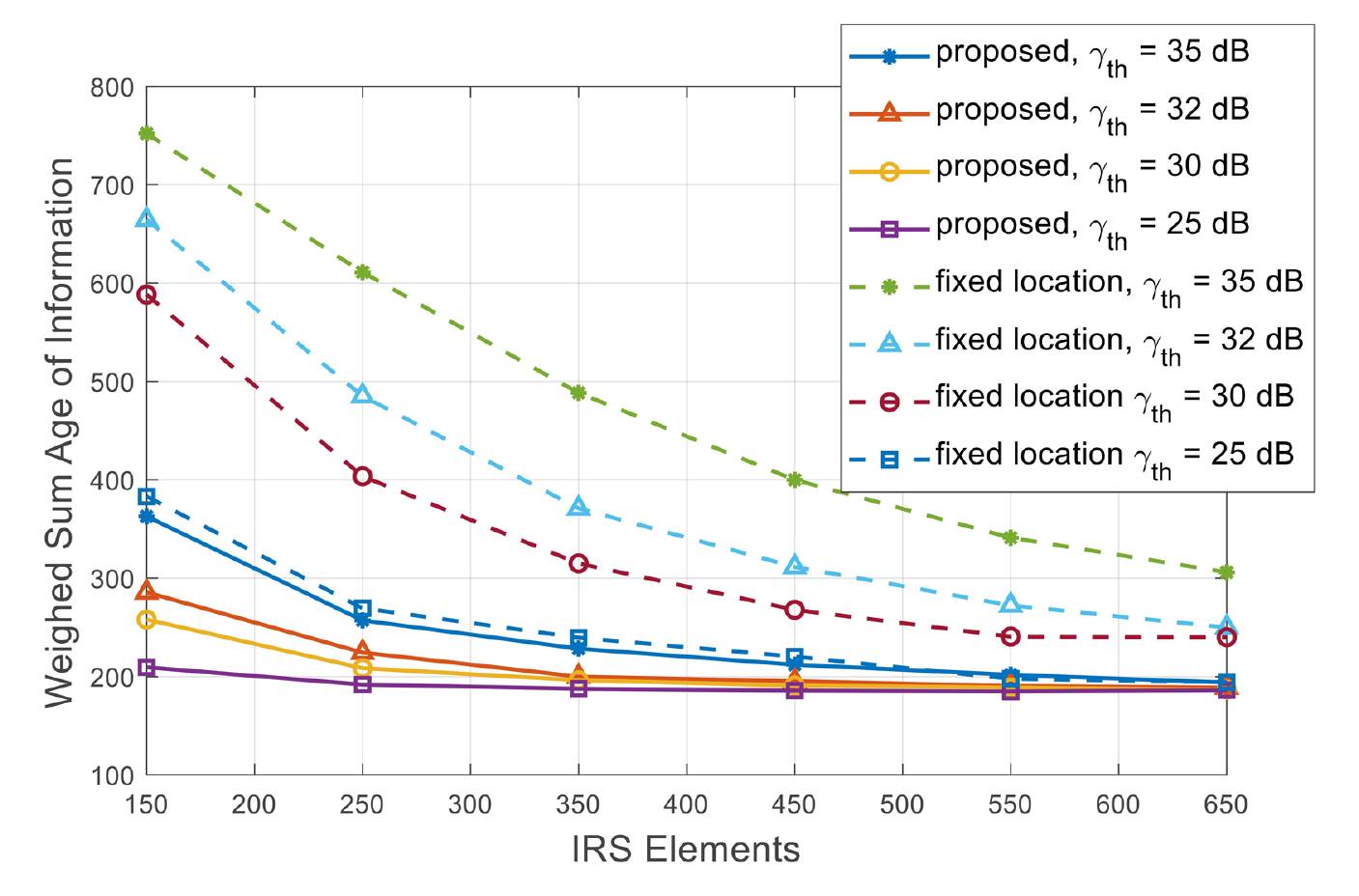}
    \caption{Weighted sum AoI versus number of IRS elements $N_s$.}
    \label{AoIvsIRS}
\end{figure}
The convergence behavior with iteration number from $2$ to $15$ is highlighted in Fig.5. We compare the AoI performance by modifying the flying height $Z$, SNR threshold $\gamma_{th}$, as well as the number of IRS elements $N_s$. Obviously, lower altitude results in better performance, due to the lower path loss. Furthermore, it can be observed that the proposed algorithm converges quickly and efficiently within about 5 iterations, verifying the low computational complexity of our proposed SCA-based solution.
\begin{figure}[!t]
    \centering
    \includegraphics[width=0.7\linewidth]{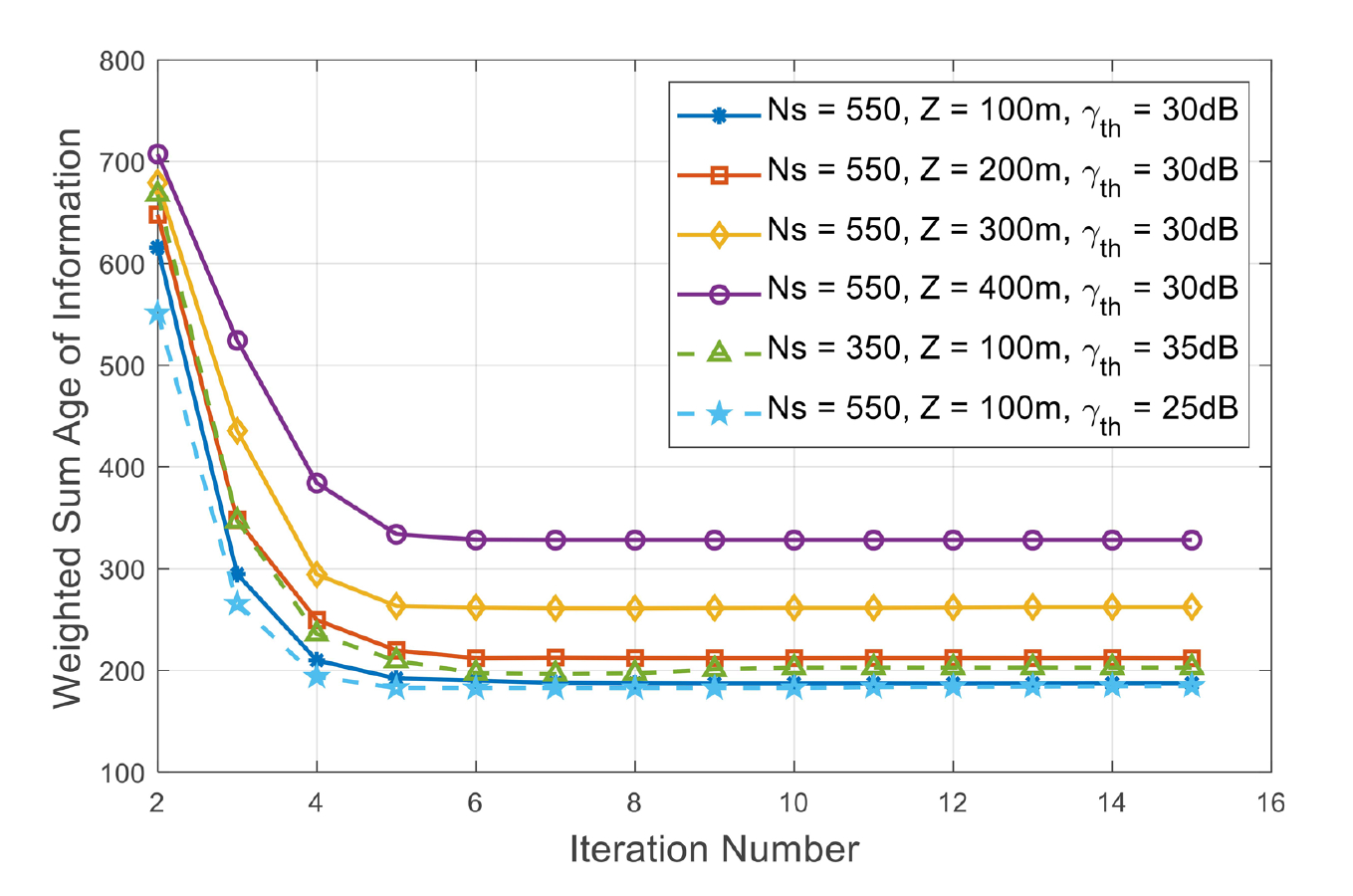}
    \caption{Convergence behavior of the proposed algorithm.}
    \label{convergence}
\end{figure}

\section{Conclusion}
We investigated the AoI performance of an aerial IRS-aided multi-user wireless network, where a user scheduling model with joint active and passive beamforming and UAV trajectory is designed to minimize the system weighted sum AoI. The SCA-based optimization algorithm is proposed to efficiently overcome the non-convexity of the formulated problem. The significant performance improvement and the convergence behavior with the proposed low complexity optimization algorithm is illustrated in the numerical results. We showed that the flexible aerial IRS deployment can significantly enhance the communication performance and information freshness for real-time network applications. 


%

\ifCLASSOPTIONcaptionsoff
  \newpage
\fi

\footnotesize
\bibliographystyle{IEEEtran}
\bibliography{Text}

\end{document}